\documentclass[a4paper,11pt]{article}

\usepackage{pos}
\usepackage{siunitx}
\usepackage{float}
\usepackage{graphicx}

\title{Search for the Prompt Atmospheric Neutrino Flux in IceCube}

\ShortTitle{Prompt Neutrino Flux Search}

\author{The IceCube Collaboration \\{\normalsize \normalfont(a complete list of authors can be found at the end of the proceedings)}\\}

\emailAdd{jakob.boettcher@rwth-aachen.de}

\abstract{

For about a decade the IceCube Neutrino Observatory has been observing a high-energy diffuse astrophysical neutrino flux. At these energies, an important source of background are the prompt atmospheric neutrinos produced in decays of charmed mesons that are part of cosmic-ray-induced air showers. The production yield of charmed mesons in the very forward phase space of hadronic interactions, and thus the flux of prompt neutrinos, is not well known and has not yet been observed by IceCube. A measurement of the flux of prompt neutrinos will improve the modeling of hadronic interactions in cosmic-ray induced air showers at high energies. Additionally, in the context of astrophysical neutrino measurements, understanding this background flux will improve the measurement precision of the spectral shape in the future. In particular, the analysis of up-going muon neutrino-induced tracks in IceCube provides a large sample of atmospheric neutrinos which likely includes prompt neutrinos. However, the measurement of a subdominant prompt neutrino flux strongly depends on the hypothesis for the dominant astrophysical neutrino flux. This makes the estimation of upper limits on the prompt neutrino flux challenging. We discuss the extent of this model dependency on the astrophysical flux and propose a method to calculate robust upper limits. Furthermore, a possible dedicated search of the prompt neutrino flux using multiple IceCube detection channels is outlined.

\vspace{4mm}
{\bfseries Corresponding authors:}
Jakob Böttcher $^{1*}$\\
{$^{1}$ \itshape RWTH Aachen}\\
$^*$ Presenter

\ConferenceLogo{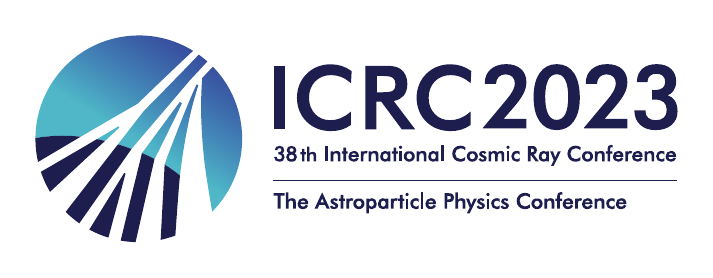}

\FullConference{The 38th International Cosmic Ray Conference (ICRC2023)\\ 26 July -- 3 August, 2023\\ Nagoya, Japan}
}

\begin{document}

\maketitle

\section{Introduction}\label{sec:Introduction}

\begin{figure}[h]
    \centering
    \includegraphics[width = 0.9\textwidth]{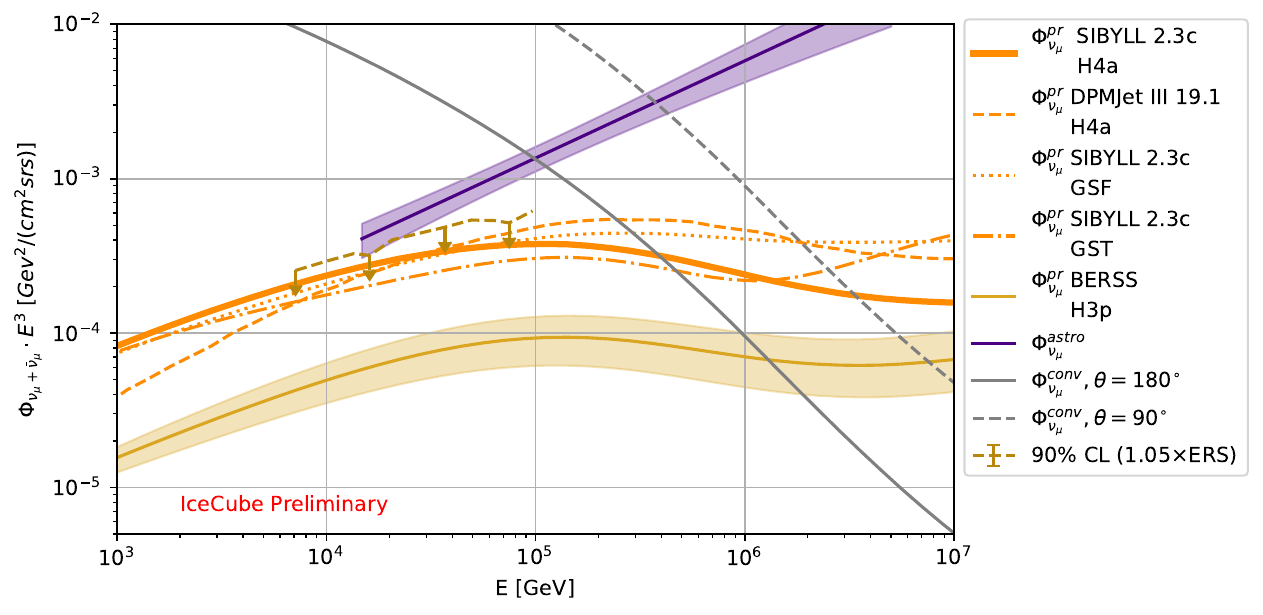}
    \caption{Comparison of different models predicting the prompt atmospheric neutrino flux. This includes SIBYLL 2.3c \cite{Riehn:2017mfm} (orange-solid), DPMJet III 19.1 \cite{Fedynitch:2018cbl} (orange-dashed) and BERSS \cite{Bhattacharya:2015jpa} (yellow-solid) as well as variations of the primary cosmic ray model (H4a \cite{Gaisser:2011klf} (orange-solid), GSF \cite{Dembinski:2017zsh} (orange-dotted), GST \cite{Gaisser:2013bla} (orange dash-dotted)). For comparison, the  conventional neutrino flux between vertical (solid) and horizontal (dashed) arrival directions is plotted in grey (based on SIBYLL 2.3c and H4a). Also the single power law description of the astrophysical neutrino flux of the latest iteration of the up-going muon tracks analysis \cite{IceCube:2021uhz} is shown for comparison in purple as well as the IceCube prompt neutrino limit from 2016 (yellow dashed) \cite{IceCube:2016umi} which is based on \cite{Enberg:2008te}. }
    \label{fig:PromptComparison}
\end{figure}

Atmospheric muon neutrinos are decay products of mesons in cosmic ray air-showers. These mesons themselves are produced in early stages of the air-shower development. The parent mesons of the so-called conventional part of the atmospheric muon neutrino flux are pions and kaons \cite{GaisserBook}. With increasing energy, kaons and pions are more likely to interact and heavy short-lived mesons, such as D$^0$ or D$^{\pm}$ become relevant. The flux of the neutrinos from these heavy meson decays is called prompt atmospheric neutrino flux. Modeling of the production cross section of the heavy mesons is challenging due to poor coverage by collider experiments in the relevant phase space \cite{Fedynitch:2022vty}. Therefore, the predicted flux of prompt atmospheric neutrinos can vary by almost an order of magnitude as can be seen in Figure \ref{fig:PromptComparison}. Additionally, the uncertainty of the composition of primary cosmic rays between the knee and ankle leads to variations as well. 

So far, the flux of prompt atmospheric neutrinos has not been measured, but it is one of the largest backgrounds in the search of astrophysical neutrinos with IceCube \cite{IceCube:2021uhz}. At the same time, uncertainties about shape of the astrophysical spectrum make measuring the prompt atmospheric neutrino flux difficult.

At first we introduce the IceCube Neutrino Observatory and its sensitivity to the prompt neutrino flux. In Section \ref{sec:Prompt_tracks} we then introduce the diffuse analysis of up-going muon tracks and our method of estimating sensitivities. 
Finally, We discuss the impact of astrophysical neutrino flux assumptions and a method to quantify potential biases in Section \ref{sec:astro_prompt}. 

\section{Prompt Atmospheric Neutrino Flux in IceCube}\label{sec:Prompt_tracks}

\subsection{Analysing the Muon Neutrino Flux of IceCube }
\begin{figure}
    \centering
    \includegraphics[width = 0.9\textwidth]{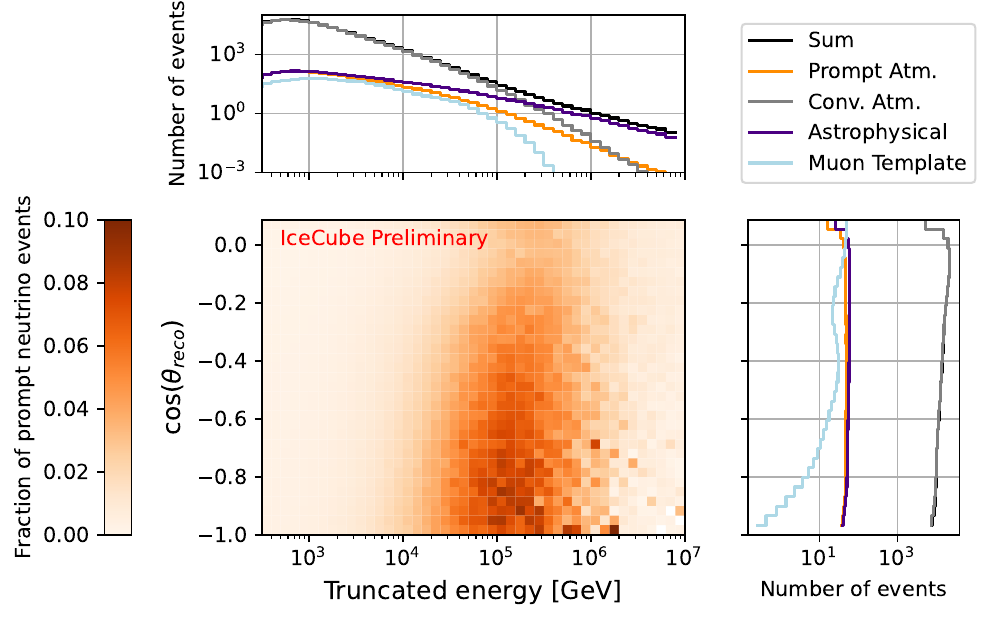}
    \caption{In the center plot, the expected fraction of prompt atmospheric neutrinos for each analysis bins of the up-going muon track analysis is shown. The plots to the top and to the right show the expected number of events for each contributing class in dependence of reconstructed energy and zenith. The assumed prompt (orange) and conventional (grey) atmospheric neutrino fluxes are based on SYBILL 2.3c model with H4a as primary flux. The astrophysical (purple) neutrino flux is modeled by a single power law. The "muon template" (light blue) models down-going atmospheric muon tracks, which were misrecontructed.  }
    \label{fig:PromptFraction}
\end{figure}
The IceCube Neutrino Observatory is a neutrino detector consisting of over 5000 photo-multiplier tubes on 86 strings embedded in the Antarctic ice \cite{Aartsen:2016nxy}. The main goal of the detector is to measure astrophysical neutrinos based on several selections of neutrino events. These selections have published multiple limits on the prompt atmospheric neutrino flux as well \cite{IceCube:2016umi,IceCube:2020acn,IceCube:2020wum}. These limits correspond to a flux of about \SI{5e-19}{cm^{-2}s^{-1}sr^{-1}GeV^{-1}} at \SI{100}{TeV}.

We use a sample of neutrino induced muon tracks \cite{IceCube:2021uhz}, to find prompt atmospheric neutrinos. The sample selects up-going tracks that originate from below the horizon, effectively excluding most atmospheric muons and achieving a purity of 99.7\%. 
In 8 years over 600,000 neutrino events were observed with most of them being atmospheric neutrinos. A large number of these observed neutrinos is expected  to be a prompt atmospheric neutrino. As most muon tracks start outside the detector volume, the energy of the primary neutrino can not be fully reconstructed. But due to the km length of the muon tracks, the angular resolution is below \SI{1}{\degree} above energies of \SI{1}{TeV}.

To analyze this sample, a parameter forward-folding fit is performed. We weight Monte-Carlo (MC) events  to a flux prediction and then compare it to data in bins of reconstructed energy and zenith. To obtain the best-fit for all the model parameters, we maximize a Poisson Log-Likelihood (LLH) based on the expected and measured event numbers in each analysis bin. The analysis binning as well as the expected number of prompt atmospheric neutrinos is depicted in Figure \ref{fig:PromptFraction}. The largest sensitivity to the prompt atmospheric neutrino flux is expected at reconstructed energies between \SI{10}{TeV} and \SI{1}{PeV} and at vertically up-going directions. The prompt atmospheric neutrino flux (modeled by SIBYLL 2.3c \cite{Riehn:2017mfm}) is altered by a normalization factor $\Phi^{0}_{\mathrm{prompt}}$. Variations in shape are considered by a $\delta\gamma$ factor and the fit interpolates between the H4a \cite{Gaisser:2011klf} and GST \cite{Gaisser:2013bla} primary cosmic ray models. These two factors model uncertainties of the primary cosmic ray flux and are applied to the conventional neutrino flux as well.

The modeling of the detector systematics is done via the "snowstorm" method \cite{IceCube:2019lxi}. By varying the detector systematics on a statistical basis, we get an ensemble of events which represent the dependency on these systematic effects. We split the ensemble into equal parts and can estimate gradients. This method saves computational effort compared to the discrete systematics used in \cite{IceCube:2021uhz}.

\subsection{Central Limit Construction}\label{sec:LimitMeasurement}

\begin{figure}
    \centering
    \includegraphics[width = 0.9\textwidth]{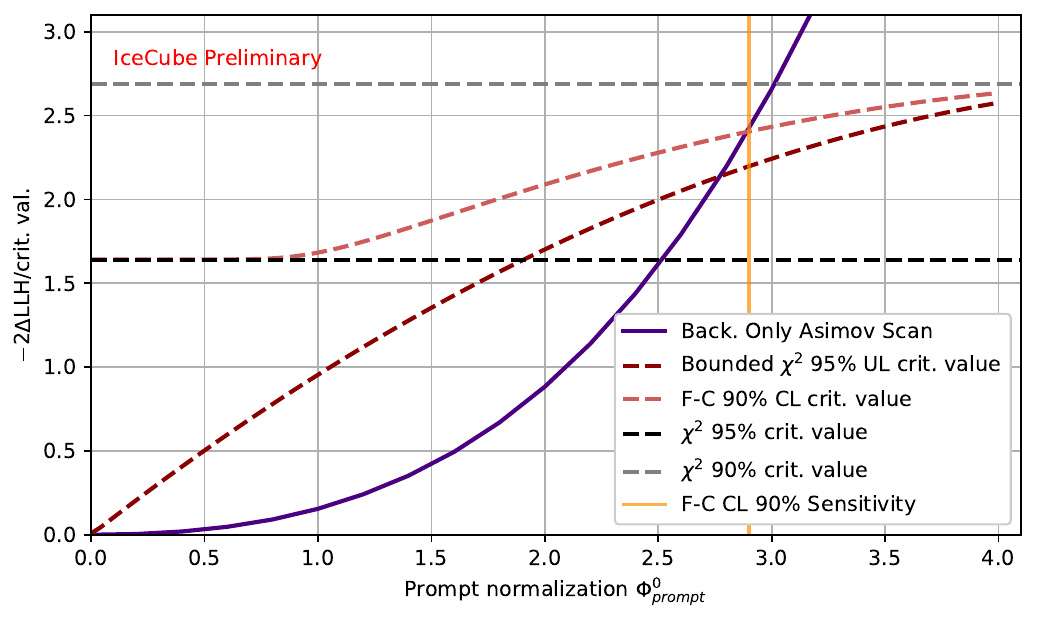}
    \caption{Visualization of the sensitivity and upper limit construction. The solid purple line shows the likelihood difference toward the background only hypothesis, setting $\Phi^0_{\mathrm{prompt}}$ to zero. As reference lines the critical value lines of several test statistic distributions are shown as well. For a $\chi^2$ distributed variable the 90\% (grey) and 95\% (black)  two sided critical values are drawn as straight lines. For a bounded $\chi^2$ distribution the one sided (upper limit) 95\% critical values (dark red) are drawn. For the Feldman-Cousins ordering the 95\% critical values are shown (light red). The crossing point between the Asimov scan and the Feldman-Cousins 90\% critical values marks the expected 90\% central limit (yellow vertical line).}
    \label{fig:SensitivityPlot}
\end{figure}

Using the MC as an Asimov dataset, the sensitivity on the prompt atmospheric neutrino flux is estimated by scanning the normalization $\Phi^{0}_{\mathrm{prompt}}$. 
Figure \ref{fig:SensitivityPlot} shows the $-2 \Delta \mathrm{LLH} = -2 (\mathrm{LLH}(\mu) - \mathrm{LLH}(\tilde{\mu}) )$ between the bestfit point $\tilde{\mu}$ and fits with $\mu = \Phi^0_{\mathrm{prompt}}$. The Asimov dataset shown in this case gives the median $-2 \Delta \mathrm{LLH}$ values. In an actual measurement, the test-statistic value $\tilde{t} = -2\Delta \mathrm{LLH}$ will fluctuate around this value depending on the true value of $\Phi^0_{\mathrm{prompt}}$. Evaluating the cumulative distribution at the measured $\tilde{t}$ value gives a probability. The 90\% upper or central limit can then be estimated by finding the $\Phi^0_{\mathrm{prompt}}$ value resulting in a probability of 90\%.

For a parameter that follows a normal distribution and is unbound, the $\tilde{t}$ distribution would correspond to a $\chi^2$ distribution independently of the assumed parameter value. If a parameter is bounded, the $\tilde{t}$ distribution has to be adjusted close to that boundary:
\begin{equation}
    \tilde{t}(\mu) = 
    \begin{cases}
        -2 (\mathrm{LLH}(\mu) - \mathrm{LLH}(\tilde{\mu}) )  & \tilde{\mu} > 0 \\
        -2 (\mathrm{LLH}(\mu) - \mathrm{LLH}(0) ) & \tilde{\mu} < 0,
    \end{cases}
\end{equation}
The cumulative distribution of $\tilde{t}$ which takes such a boundary into account is given by \cite{Cowan:2010js}:
\begin{equation}
    F\left(\tilde{t}_\mu \mid \mu\right)= 
    \begin{cases}
    \mathcal{G}\left(\sqrt{\tilde{t}_\mu}\right) 
    & 
    \tilde{t}_\mu \leq \mu^2 / \sigma^2 \\ 
    \mathcal{G}\left(\frac{\tilde{t}_\mu+\mu^2 / \sigma^2}{2 \mu / \sigma}\right) 
    & \tilde{t}_\mu>\mu^2 / \sigma^2,
    \end{cases}
\end{equation}

with $\mathcal{G}(x)$ being the cumulative normal distribution function and $\sigma$ the expected uncertainty on the parameter $\mu$, which can be estimated from inverting the hessian matrix of the negative likelihood function (Fisher Information). This boundary excludes any upward fluctuations as it is defined for pure one sided upper limits. This can lead to empty intervals in case of a downward fluctuation of the parameter. To avoid such a behaviour the central limit construction by Feldman and Cousins \cite{Feldman:1997qc} can be used instead. The cumulative test-statistic distribution would then be given by \cite{Cowan:2010js}:
\begin{equation}
    F\left(\tilde{t}_\mu \mid \mu\right)= 
    \begin{cases}
    2 \; \mathcal{G}\left(\sqrt{\tilde{t}_\mu}\right)-1 & 
    \tilde{t}_\mu \leq \mu^2 / \sigma^2 \\ 
    \mathcal{G}\left(\sqrt{\tilde{t}_\mu}\right)+
    \mathcal{G}\left(\frac{\tilde{t}_\mu+\mu^2 / \sigma^2}{2 \mu / \sigma}\right)-1 &
    \tilde{t}_\mu>\mu^2 / \sigma^2.
    \end{cases}
\end{equation}
The comparison of the 90\% intervals depending on the assumed true normalization value is shown in Figure \ref{fig:SensitivityPlot}. The expected 90\% sensitivity is given by the point the background only scan of $\Phi^0_{\mathrm{prompt}}$ reaches the critical value given by the 90\% lines of the different intervals. The large difference between the upper limit construction and the Feldman-Cousins central limit stems from the effect that the first is a purely one sided test and the second is two sided distribution which can be used for lower limits as well. In both cases this depends strongly on the assumption of the astrophysical model. We will discuss this dependency in the next section.

\section{Consequences of the Assumption of an Astrophysical Neutrino Flux Model}\label{sec:astro_prompt}
\begin{figure}
    \centering
    \includegraphics[width=0.85\textwidth]{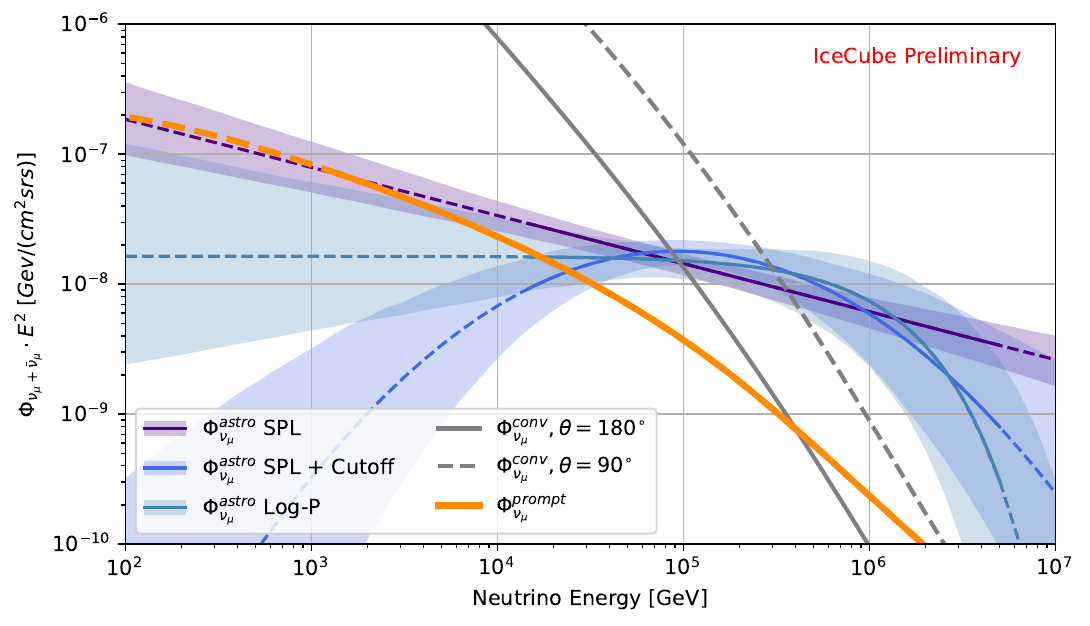}
    \caption{Plot comparing different astrophysical neutrino flux models (blue lines) to the prompt atmospheric neutrino flux (orange). Each of the astrophysical models is varied according to the uncertainties of the published result (see Table \ref{tab:astromodels}). The solid part of the lines mark the sensitive energy range (estimated for the single powerlaw model assumption), while the dashed part marks the extrapolated region. The grey lines give the conventional atmospheric neutrino flux between vertical (solid) and horizontal (dashed) arrival directions. The prompt and conventional fluxes have no uncertainties included in this plot.}
    \label{fig:Astro_modelVariation}
\end{figure}

\begin{table}
\caption{List of the different astrophysical models used in this analysis. The normalisation factor $C_0$ in the function definitions is set to $C_0 = 10^{-18}$/(cm$^2$ s sr GeV). The factor $\chi(E)$ describes the borders of each individual piece and is defined in \cite{IceCube:2020acn}. The assumed uncertainties do not take correlations into account and serve as upper boundaries of the true uncertainty. The sensitivities are given in terms of the prompt normalization factor $\Phi^0_{\mathrm{prompt}}$ and are estimated using the F-C method introduced in Section \ref{sec:LimitMeasurement}.}\label{tab:astromodels}
\begin{tabular}{p{2.2cm}|p{5.1cm}|p{4cm}|p{1.5cm}}

Astrophysical Model &  Function   &Assumed Values &  <$\Phi^0_{\mathrm{prompt}}$>\newline 90\%-CL\\ \hline
Powerlaw            &  $\Phi^0_{\mathrm{astro}}/C_0 (\frac{E}{\SI{100}{TeV}})^{-\gamma}$  & $\Phi^0_{\mathrm{astro}} = 1.44\pm0.26$,\newline $\gamma = 2.37\pm0.09$   & 2.9  \\
Log-Parabola          & $\Phi^0_{\mathrm{astro}}/C_0(\frac{E}{\SI{100}{TeV}})^{-\alpha-\beta\log_{10}(\frac{E}{\SI{100}{TeV}})}$ &  $\Phi^0_{\mathrm{astro}} = 1.79\pm0.4$,\newline $\alpha = 2.03\pm0.22$, \newline $\beta = 0.45\pm0.29$   & 2.45  \\
Piecewise           &   $\sum_i^5 \chi(E)\; \Phi^0_{i}/C_0\; (\frac{E}{\SI{100}{TeV}})^{-2}$       & Piecewise fit results in \cite{IceCube:2021uhz} &  2.4   \\
Cascades            &  $\sum_i^{13}  \chi(E) \; \Phi^0_{i}/C_0 \; (\frac{E}{\SI{100}{TeV}})^{-2}$ & Piecewise fit results as prior \cite{IceCube:2020acn}   &  1.45  \\
Cutoff              &  $\Phi^0_{\mathrm{astro}}/C_0(\frac{E}{\SI{100}{TeV}})^{-\gamma} \; e^{-\frac{E}{E_{\mathrm{cutoff}}}}$ & $\Phi^0_{\mathrm{astro}} = 1.64 \pm 0.39$,\newline $\gamma = 2.0\pm0.4$, \newline $\log_{10}(E_{\mathrm{cutoff}}) = 6.1\pm0.3$  &     \\
Astro BL-Lac         &  $\Phi_{\mathrm{Cutoff}}(E) + \Phi_{\mathrm{BL-Lac}}(E)$  &  Adding the BL-Lac astrophysical model \cite{10.1093/mnras/stv1467, IceCube:2020acn}& \\   
\end{tabular}
\end{table}

Previous IceCube analyses \cite{IceCube:2021uhz,IceCube:2020acn} show hints of structure in the astrophysical neutrino spectrum deviating from the single powerlaw. These mostly refer to a cutoff above energies of \SI{1}{PeV}, but these results impact the spectrum below \SI{10}{TeV} as well. This can be seen in Figure \ref{fig:Astro_modelVariation}, where the results of different spectral assumption are compared to each other for the analysis of the up-going muon tracks \cite{IceCube:2021uhz}. The definitions of these spectral assumptions are shown in Table \ref{tab:astromodels}. 
The models agree in the sensitive energy range between \SI{15}{TeV} and \SI{5}{PeV}, but outside of the sensitive energy range the astrophysical flux varies between the assumed models by orders of magnitude. At the same time the sensitive energy range on the prompt atmospheric flux extends to \SI{300}{GeV}. In this energy range the astrophysical flux has either similar strength or is well below the prompt atmospheric neutrino flux depending on the assumed astrophysical model. Therefore, we expect the resulting prompt normalization to be biased by the choice of astrophysical model, as the true shape of the astrophysical neutrino flux is not well known below \SI{10}{TeV}.  

To test the strength of this expected bias, an Asimov-like test can be used. As the true shape of the astrophysical spectrum is unknown, several model assumptions are injected and varied inside their measured uncertainties (see Table \ref{tab:astromodels}). A selection of astrophysical models are fitted to these assumptions: 
a single powerlaw, a log parabola, five spectral pieces and finally 13 spectral pieces with priors from the analysis of cascade like events in IceCube \cite{IceCube:2020acn}.
To test the full extend of the bias on $\Phi^0_{\mathrm{prompt}}$ the boundary at zero is lifted and the default prompt model is injected. The rest of the nuisance parameters are injected at their default values.

\begin{minipage}{0.45\textwidth}
\begin{figure}[H]
    \includegraphics[width = 1.0\textwidth]{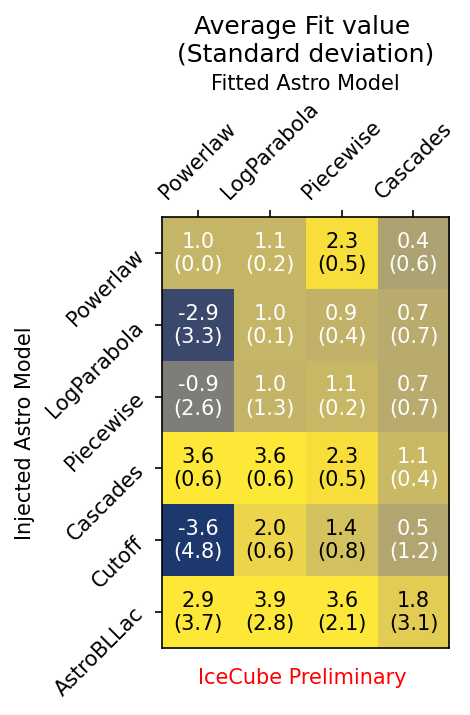}
    \caption{The bias on the prompt neutrino flux normalization for variations of the injected and fitted astrophysical neutrino flux. Each of the astrophysical neutrino flux models is varied according to uncertainties of previous fits. The median and standard deviation of the fitted (without bound at 0) prompt normalization are given.}
    \label{fig:Confusion_matrix}
    \end{figure}
\end{minipage}\hfill
\begin{minipage}{0.45\textwidth}
The result of this test can be seen in Figure \ref{fig:Confusion_matrix}, with the resulting average and standard deviation of the fitted $\Phi^0_{\mathrm{prompt}}$. In the case of fitting the same model as injected the bias becomes small as well as the variation. The single powerlaw shows the largest variation towards negative values (if different flux assumptions are injected), while the log-parabola and spectral pieces fits have smaller biases.
These astrophysical model assumptions have a varying number of fitted parameters. Therefore, such a behaviour would be expected, but it highlights the shortcomings of a single powerlaw assumption for the analysis of the prompt neutrino flux. Additionally, these models are expected to impact the sensitivity on the prompt neutrino flux.  
Using the methods introduced in Section \ref{sec:LimitMeasurement}, we can estimate this sensitivity by calculating the Fisher information and scanning the $\Phi^0_{\mathrm{prompt}}$ parameter. The resulting 90\% sensitivities are listed in Table \ref{tab:astromodels}. The cascade pieces improve the sensitivity on the prompt atmospheric neutrino flux substantially, while the other model assumptions have a comparable sensitivity.  
\end{minipage}

\section{Discussion and Outlook}
The results presented here show that the observation of the prompt atmospheric neutrino flux remains challenging using only the up-going muon track analysis. The example of IceCube's up-going muon track analysis shows that the assumption of a specific astrophysical model introduces a bias on the measurement of the prompt astrophysical flux. This dependency makes the construction of upper limits from past measurements unreliable. It can also explain the consistent under-fluctuation of the measured $\Phi^0_{\mathrm{prompt}}$ in several analyses published by IceCube. For further investigations potential biases due to conventional atmospheric flux assumptions (e.g. primary flux models) have to be investigated as well. The result of the cascade fit with prior shows the potential of the global fit effort to combine the up-going tracks analysis with the cascades. In order to measure the prompt atmospheric flux with a sufficient significance other analyses will have to be included as well.

\bibliographystyle{ICRC}
\bibliography{references}

%

\clearpage

\section*{Full Author List: IceCube Collaboration}

\scriptsize
\noindent
R. Abbasi$^{17}$,
M. Ackermann$^{63}$,
J. Adams$^{18}$,
S. K. Agarwalla$^{40,\: 64}$,
J. A. Aguilar$^{12}$,
M. Ahlers$^{22}$,
J.M. Alameddine$^{23}$,
N. M. Amin$^{44}$,
K. Andeen$^{42}$,
G. Anton$^{26}$
% ...
\\
\\
% ...
$^{12}$ Universit{\'e} Libre de Bruxelles, Science Faculty CP230, B-1050 Brussels, Belgium \\
% ...
$^{17}$ Department of Physics, Loyola University Chicago, Chicago, IL 60660, USA \\
$^{18}$ Dept. of Physics and Astronomy, University of Canterbury, Private Bag 4800, Christchurch, New Zealand \\
% ...
$^{22}$ Niels Bohr Institute, University of Copenhagen, DK-2100 Copenhagen, Denmark \\
$^{23}$ Dept. of Physics, TU Dortmund University, D-44221 Dortmund, Germany \\
% ...
$^{26}$ Erlangen Centre for Astroparticle Physics, Friedrich-Alexander-Universit{\"a}t Erlangen-N{\"u}rnberg, D-91058 Erlangen, Germany \\
% ...
$^{40}$ Dept. of Physics and Wisconsin IceCube Particle Astrophysics Center, University of Wisconsin{\textendash}Madison, Madison, WI 53706, USA \\
% ...
$^{42}$ Department of Physics, Marquette University, Milwaukee, WI, 53201, USA \\
% ...
$^{44}$ Bartol Research Institute and Dept. of Physics and Astronomy, University of Delaware, Newark, DE 19716, USA \\
% ...
$^{63}$ Deutsches Elektronen-Synchrotron DESY, Platanenallee 6, 15738 Zeuthen, Germany  \\
$^{64}$ Institute of Physics, Sachivalaya Marg, Sainik School Post, Bhubaneswar 751005, India \\
% ...

\subsection*{Acknowledgements}

\noindent
The authors gratefully acknowledge the support from the following agencies and institutions:
USA {\textendash} U.S. National Science Foundation-Office of Polar Programs,
U.S. National Science Foundation-Physics Division,
U.S. National Science Foundation-EPSCoR,
Wisconsin Alumni Research Foundation,
Center for High Throughput Computing (CHTC) at the University of Wisconsin{\textendash}Madison,
Open Science Grid (OSG),
Advanced Cyberinfrastructure Coordination Ecosystem: Services {\&} Support (ACCESS),
Frontera computing project at the Texas Advanced Computing Center,
U.S. Department of Energy-National Energy Research Scientific Computing Center,
Particle astrophysics research computing center at the University of Maryland,
Institute for Cyber-Enabled Research at Michigan State University,
and Astroparticle physics computational facility at Marquette University;
Belgium {\textendash} Funds for Scientific Research (FRS-FNRS and FWO),
FWO Odysseus and Big Science programmes,
and Belgian Federal Science Policy Office (Belspo);
Germany {\textendash} Bundesministerium f{\"u}r Bildung und Forschung (BMBF),
Deutsche Forschungsgemeinschaft (DFG),
Helmholtz Alliance for Astroparticle Physics (HAP),
Initiative and Networking Fund of the Helmholtz Association,
Deutsches Elektronen Synchrotron (DESY),
and High Performance Computing cluster of the RWTH Aachen;
Sweden {\textendash} Swedish Research Council,
Swedish Polar Research Secretariat,
Swedish National Infrastructure for Computing (SNIC),
and Knut and Alice Wallenberg Foundation;
European Union {\textendash} EGI Advanced Computing for research;
Australia {\textendash} Australian Research Council;
Canada {\textendash} Natural Sciences and Engineering Research Council of Canada,
Calcul Qu{\'e}bec, Compute Ontario, Canada Foundation for Innovation, WestGrid, and Compute Canada;
Denmark {\textendash} Villum Fonden, Carlsberg Foundation, and European Commission;
New Zealand {\textendash} Marsden Fund;
Japan {\textendash} Japan Society for Promotion of Science (JSPS)
and Institute for Global Prominent Research (IGPR) of Chiba University;
Korea {\textendash} National Research Foundation of Korea (NRF);
Switzerland {\textendash} Swiss National Science Foundation (SNSF);
United Kingdom {\textendash} Department of Physics, University of Oxford.

\end{document}